\documentclass[amsmath,11pt,amssymb]{revtex4}

\usepackage{color}
\usepackage[]{graphicx}
\usepackage{latexsym}
\usepackage{natbib}
\usepackage{amsmath}
\usepackage[caption=false]{subfig}
\usepackage{multirow}
\usepackage{mathtools}
\usepackage{textcomp}
\usepackage{xr}

\usepackage[pdftex,hyperfigures,breaklinks,colorlinks,citecolor=black,linkcolor=black]{hyperref} 

\newcommand{\be}{\begin{equation}}
\newcommand{\ee}{\end{equation}}
\newcommand{\bc}{\begin{cases}}
\newcommand{\ec}{\end{cases}}

\renewcommand{\figurename}{Fig.}

\begin{document}

\title{Clustering matrices through optimal permutations}
\author{Flaviano Morone}
\affiliation{Department of Radiology, Memorial Sloan Kettering 
Cancer Center, New York, NY, 10065, USA}

\begin{abstract}
Matrices are two-dimensional data structures allowing one 
to conceptually organize information~\cite{golub}. 
For example, adjacency matrices are useful to store the links 
of a network; correlation matrices are simple ways to arrange gene 
co-expression data or correlations of neuronal activities~\cite{patrik,zimmer}. 
Clustering matrix values into geometric patterns that are easy 
to interpret~\cite{jain} helps us to understand and explain 
the functional and structural organization of the system 
components described by matrix entries. 
Here we introduce a theoretical framework to cluster a matrix into 
a desired pattern by performing a similarity transformation obtained 
by solving a minimization problem named the optimal permutation problem. 
On the computational side, we present a fast clustering algorithm 
that can be applied to any type of matrix, including non-normal and 
singular matrices. 
We apply our algorithm to the neuronal correlation matrix and 
the synaptic adjacency matrix of the {\it Caenorhabditis elegans} 
nervous system by performing different types of clustering, 
including block-diagonal, nested, banded, and triangular 
patterns. Some of these clustering patterns show their biological 
significance in that they separate matrix entries into groups that match 
the experimentally known classification of {\it C. elegans} neurons into 
four broad categories, namely: interneurons, motor, sensory, and polymodal 
neurons.  

\end{abstract}

\maketitle

\section{Introduction}
We formulate the optimal permutation problem (OPP) in a 
pragmatic way by considering the correlation matrix $B$ showed in 
Figure~\ref{fig:figure_1}a describing the neuronal activity of $N = 33$ 
neurons of the nematode {\it C. elegans} measured for three locomotory 
tasks of the animal (forward, backward, turn) in ref.~\cite{zimmer}. 
The choice of this particular dataset is useful to get an instrumental 
view of the optimal permutation problem and how it relates to real-world 
data, although we could formulate our mathematical theory completely  
{\it in abstracto} as well. Thus, we emphasize that the optimal 
permutation theory and algorithm we present here can be applied to 
any square matrix of the most general form.

The entries $B_{ij}$ of the correlation matrix in Figure~\ref{fig:figure_1}a 
take values in the interval $B_{ij}\in[-1,1]$, where the extreme value $B_{ij}=1$ 
occurs for neurons $i$ and $j$ which are both active during locomotion. 
The other extreme value $B_{ij}=-1$, instead, occurs whenever neuron $i$ 
is active while neuron $j$ is not, and viceversa.
The existence of positive and negative correlations implies the existence 
of at least two groups of neurons such that all neurons in one group are 
positively correlated with each other and negatively correlated with neurons 
in the other group. The presence of two groups of neurons can be 
traced back in the twofold nature of the locomotion behavior comprising: 
forward movement mediated by one group of neurons, and backward 
movement mediated by another one (reversal and turns 
are accounted for by a third group of neurons, as we explain at the end of this 
section).

To identify the two groups, we consider a matrix $A$, called filter, with a 
two-blocks shape, as seen in Figure~\ref{fig:figure_1}b. Precisely, $A_{ij}=1$ 
if $i,j\in [1,N/2]\times[1,N/2]\cup[1+N/2,N]\times[1+N/2,N]$, and $A_{ij}=0$ 
otherwise. 
The role of the filter matrix $A$ is to conceptualize visually how the matrix 
$B$ ought to be clustered into two blocks. In other words, we use $A$ to 
guide the clustering process in order to get a clustered matrix $B'$ as 
`similar' as possible to $A$. 
Mathematically, this can be achieved by means of the objective function 
$E(P)$ defined as 
\be
E(P) = ||PA-BP||^2 = ||A||^2 + ||B||^2 -2{\rm tr}(B^{\rm t}PAP^{\rm t})
\label{eq:costfunction}
\ee
where $||A||^2={\rm tr}(A^{\rm t}A)$ is the Frobenius norm and 
$P$ is a permutation matrix, whose entries $P_{ij}\in\{0,1\}$ satisfy 
the constraints $\sum_iP_{ij}=\sum_jP_{ij}=1$. The objective function in 
Equation~\eqref{eq:costfunction} appeared for the first time 
in the formulation of the quadratic assignment problem (QAP)~\cite{koopmans}, 
which is one of the most important problems in the field of combinatorial 
optimization. 
The QAP was initially introduced in economics to find the optimal 
assignement (=optimal permutation in our language) of $N$ facilities 
to $N$ locations, given the matrix of distances between pair of facilities 
(=filter matrix $A$) and a weight matrix quantifying the amount of goods 
flowing between firms (=correlation matrix $B$). 

To better explain the meaning of the objective function in Equation~\eqref{eq:costfunction}, 
let us suppose that we could find a permutation matrix $P_*$ such that $E(P_*)=0$. 
This means that matrix $A$ is permutation similar to matrix $B$ via 
the transformation $A=P_*^{\rm t}BP_*$. That is, $A$ itself is the desired clustering 
of the matrix $B$. 
But the equation $E(P)=0$ has no solution almost always (it admits solutions only 
for special choices of the filter $A$). Therefore, $A$ is not permutation similar to 
$B$ and it's not itself a clustering of $B$. 
What we can do in this situation is to look for a permutation matrix 
$P_*$ that minimizes the cost function so that the weaker condition 
$E(P_*)\geq 0$ holds true, as seen in Figure~\ref{fig:figure_1}d, that is 
the solution of the following optimization 
problem:
\be 
%P_* = {\rm arg}\min_{P\in\mathcal{P}}E(P)\ .
P_* = {\rm arg}\min_P E(P)\ .
\label{eq:optimalpermutation}
\ee 
We call $P_*$ the {\bf optimal permutation} of matrix $B$ 
(given the filter matrix $A$) and we show it in Figure~\ref{fig:figure_1}e. 
Once we obtain $P_*$ we can proceed to cluster matrix $B$ by performing 
a similarity transformation to bring $B$ into its clustered form $B'$:
\be
B' = P_*^{\rm t}BP_*\ .
\label{eq:clusteredmatrix}
\ee 
The result is shown in Figure~\ref{fig:figure_1}f. The two-blocks clustering (left 
panel in Figure~\ref{fig:figure_1}f) identifies two clusters separating two groups 
of neurons: one group contains neurons driving backward locomotion, and the 
other one contains those regulating forward locomotion.
By using the three-blocks filter shown in Figure~\ref{fig:figure_1}c we obtain 
a clustering of the correlation matrix $B$ into $3$ clusters: two of them are 
each a subset of the backward and forward locomotion groups defined previously. 
The third cluster occupies the middle block in the right panel of 
Figure~\ref{fig:figure_1}f. Neurons belonging to this cluster are classified 
by the Wormatlas database~\cite{wormatlas} as: ring interneurons (RIVL/RIVR) 
regulating reversals and deep omega-shaped turns; motor neurons 
(SMDVR/SMDVL, RMEV) defining the amplitude of omega turns; 
labial neurons (OLQDR/OLQVL) regulating nose oscillations in local search 
behavior; and a high-threshold mechanosensor (ALA) responding to harsh-touch 
mechanical stimuli. 
We term ``Turn'' the third block in the clustered correlation matrix shown in 
Figure~\ref{fig:figure_1}f. 

Having formulated the optimal permutation problem, we move now to 
explain the algorithm to solve it, along with several interesting applications 
to the {\it C. elegans} whole brain's connectome.

\section{Solution to the Optimal Permutation Problem}
In order to determine the solution to the optimal permutation 
problem (OPP) given in Equation~\eqref{eq:optimalpermutation} 
we use Statistical Mechanics methods~\cite{zinn}. 
The quantity which plays the fundamental role in the resolution 
of the OPP is the partition function $Z(\beta)$, defined as 
\be
Z(\beta) = \sum_{P} e^{-\beta E(P)}\ ,
\label{eq:partitionfunction}
\ee
where the sum is over all permutation matrices $P$. 
The statistical physics interpretation of the problem thus follows.
The parameter $\beta$ in Equation~\eqref{eq:partitionfunction} represents
the inverse of the `temperature' of the system; the cost function $E(P)$ defined 
in Equation~\eqref{eq:costfunction} becomes the `energy' function. 
The global minimum of the energy function corresponds to the `ground-state' 
of the system. Since a physical system goes into its ground state only at zero 
temperature (by the third law of thermodynamics), then the exact solution to 
the OPP corresponds to the zero temperature 
limit of the partitition function in Equation~\eqref{eq:partitionfunction}: 
\be
\lim_{\beta\to\infty}-\frac{1}{\beta}\log Z(\beta) 
= \min_{P}E(P) = E(P_*)\ . 
\label{eq:zerotemperature}
\ee 
In this limit the partition function can be evaluated by the steepest 
descent method, which leads us to the following saddle point equations 
(see SI Sec.~\ref{sec:theory} for the detailed calculations):
\be
X_{ij} = U_i(X)Y_{ij}(X)V_j(X)\ ,
\label{eq:saddlepoint}
\ee
where $X_{ij}$ are the entries of a double-stochastic matrix $X$, 
that take values in the interval $X_{ij}\in[0,1]$ and satisfy normalization 
conditions on row and column sums: $\sum_iX_{ij}=\sum_jX_{ij}=1$ for 
all $i$ and $j$ (the space of all $X$'s is also called the Birkhoff 
polytope~\cite{linderman}).
The matrix $Y(X)$ contains the information about matrices $A$ 
and $B$ and its components are explicitely given by
\be
Y_{ij}(X) = \exp\left[\frac{\beta}{2}\Big(BXA^{\rm t}+B^{\rm t}XA\Big)_{ij}\right]\ .
\label{eq:ymatrix}
\ee
The vectors $U$ and $V$ are needed to ensure the row and column 
normalization conditions, and we compute them by solving the 
Sinkhorn-Knopp equations~\cite{sinkhorn1, sinkhorn2, cuturi}:
\be 
\begin{aligned}
U_i^{-1}(X) &=  \sum_{j=1}^NY_{ij}(X) V_j(X)\ ,\\
V_j^{-1}(X) &= \sum_{i=1}^N U_i(X) Y_{ij}(X)\ .
\end{aligned}
\label{eq:sinkhorn}
\ee
Equations~\eqref{eq:saddlepoint} and~\eqref{eq:ymatrix}, 
represent our main result. Equations similar to~\eqref{eq:sinkhorn} 
have been already derived in ref.~\cite{tacchella} to relate 
the fitness of countries to their economic complexity. 
Note that the solution $X_*$ to the saddle point equations~\eqref{eq:saddlepoint} 
is not a permutation matrix for $\beta<\infty$. To find the optimal 
permutation matrix $P_*$ defined in Equation~\eqref{eq:optimalpermutation} 
we have to take the zero temperature limit by sending $\beta\to\infty$: 
in this limit the solution matrix $X_*(\beta)$ is projected onto one of 
the $N!$ vertices of the Birkhoff polytope: 
\be
\lim_{\beta\to\infty}X_*(\beta) = P_*\ ,
\ee
which is the optimal permutation matrix $P_*$ that solves the 
OPP~\cite{benzi} (details in Section~\ref{sec:theory}). 
The implementation of the algorithm to solve the saddle point 
equations~\eqref{eq:saddlepoint} is described in detail in SI Sec.~\ref{sec:algorithm}. 
Next, we use our optimal permutation algorithm to perform three 
types of clustering of the {\it C. elegans} connectome.

\section{Clustering the C.elegans connectome through optimal permutations}
\begin{figure}[h]
\includegraphics[width=0.95\textwidth]{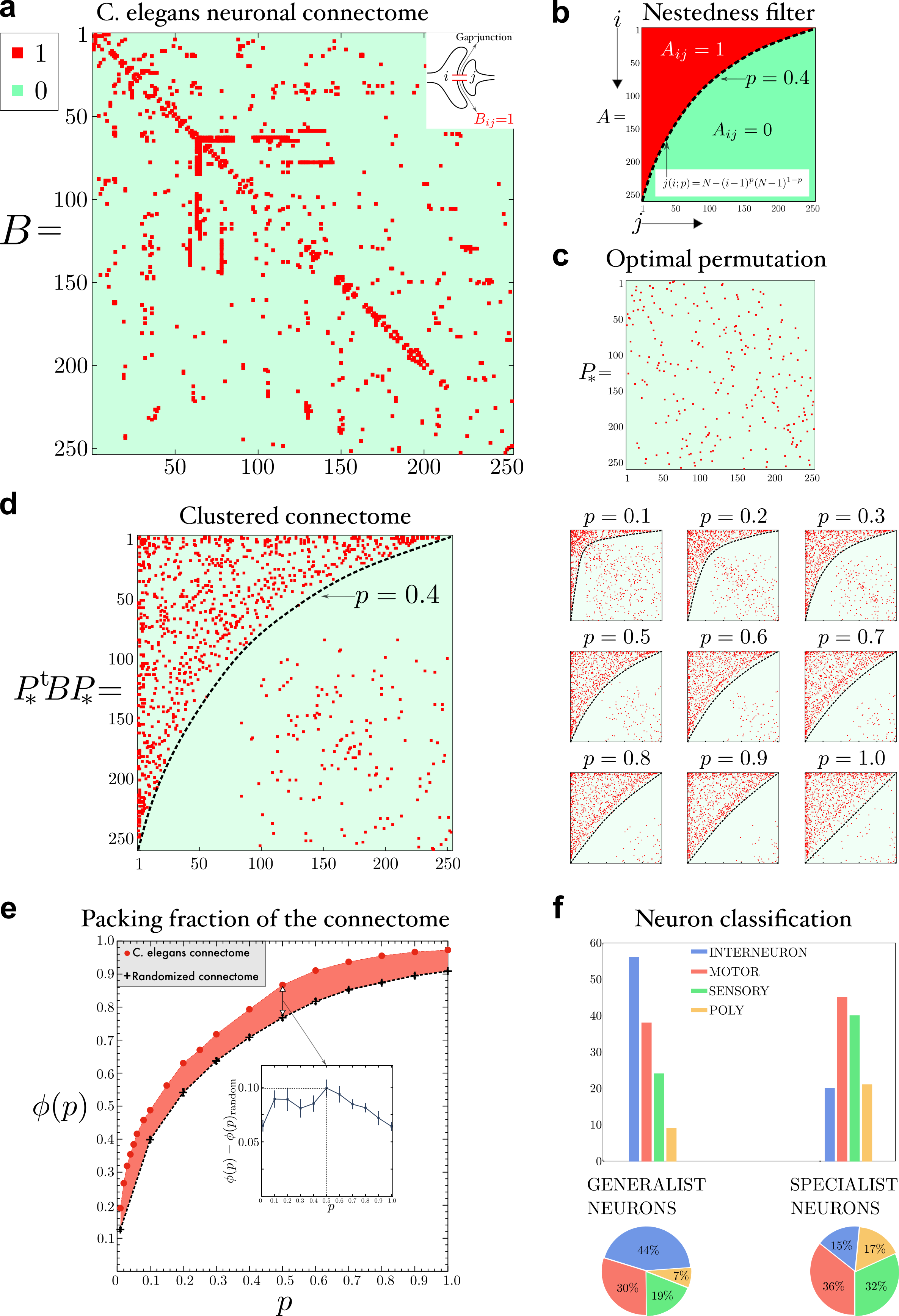} 
\vspace*{-2mm}
\caption{}
\label{fig:figure_2} 
\end{figure}

We consider the neuronal network of the hermaphrodite
{\it C. elegans} comprising $N = 253$ neurons interconnected via %$1028$ 
gap junctions (we consider only the giant component of this network). 
We use the most up-to-date connectome of gap-junctions from Ref.~\cite{varshney}. 
We represent the synaptic connectivity structure via a binary adjacency 
matrix $B$, with $B_{ij}=1$ if neuron $i$ connects (i.e. form gap-junctions) 
to $j$, and $B_{ij}=0$ otherwise, as shown in Figure~\ref{fig:figure_2}a 
($\sum_{ij}B_{ij} = 2M = 1028$, so that the mean degree is $\langle k\rangle = 2M/N \sim 4$). 
Gap-junctions are undirected links, hence $B$ is a symmetric matrix. 
We emphasize that our framework is not limited to symmetric matrices 
and can be equally applied to asymmetric adjacency matrices representing 
directed chemical synapses. 

We perform a clustering experiment by using a filter $A$ whose 
shape is shown in Figure~\ref{fig:figure_2}b. We call it: `nestedness filter'. 
This nomenclature is motivated by ecological studies of species abundance 
showing nested patterns in the community structure of mammals~\cite{patterson} 
and plant-pollinator ecosystems~\cite{bascompte, staniczenko}. Nestedness 
is also found in interbank, communication, and socio-economic 
networks~\cite{tacchella, tessone, mariani}. Remarkably, it has been shown 
recently that behavioral control in {\it C. elegans} occurs via a nested neuronal 
dynamics across motor circuits~\cite{kaplan}. 
A connectivity structure which is nested implies the existence of two types 
of nodes, either animal species, neurons or firms, that are called `generalists' 
and `specialists'. 
Generalists are ubiquitous species with a large number of links to other species 
that are quickly reachable by the other nodes; specialists are rare species with 
a small number of connections occupying peripheral locations of the network 
and having a higher likelihood to go extict~\cite{moronekcore}. 

The entries of $A$ are defined by:
\be
\begin{aligned}
&A_{ij} = 1\ \ \ {\rm if}\ \ \ j \leq f(i;p) = N - (i-1)^p(N-1)^{1-p}\\
&A_{ij} = 0\ \ \ {\rm otherwise}
\end{aligned}
\ ,
\label{nestedfilter}
\ee
where $p$ is the nestedness exponent controlling the curvature of 
the function $f(i;p)$ separating the filled and empty parts of the 
matrix $A$, as seen in Figure~\ref{fig:figure_2}b. 
By solving the OPP we obtain the optimal permutation matrix $P_*$ 
shown in Figure~\ref{fig:figure_2}c, by means of which we cluster the 
adjacency matrix via the similarity transformation $B'=P_*^{\rm t}BP_*$, 
as depicted in Figure~\ref{fig:figure_2}d. 
In order to measure the degree of nestedness of the connectome we 
introduce the quantity $\phi(p)$, defined as the fraction of elements 
of $B'$ comprised in the nested region $j\leq f(i;p)$ by the following 
formula:
\be
\phi(p) =\frac{ \sum_{i=1}^N \sum_{j=1}^{f(i; p)}(P_*^{\rm t}BP_*)_{ij}}
{ \sum_{i=1}^N \sum_{j=1}^{N}(P_*^{\rm t}BP_*)_{ij}} = 
\frac{1}{2M} \sum_{i=1}^N \sum_{j=1}^{f(i; p)}(P_*^{\rm t}BP_*)_{ij}\ .
\label{eq:packingfraction}
\ee
We call $\phi(p)$ the `packing fraction' of the network. The profile 
of $\phi(p)$ as a function of $p$ is shown in Figure~\ref{fig:figure_2}e, 
comparing the {\it C.elegans} connectome to a randomized connectome 
having the same degree sequence but neurons wired at 
random through the configurational model~\cite{newman}. 
Figure~\ref{fig:figure_2}e shows that the  {\it C.elegans}  connectome is $10\%$ 
more packed than its random counterpart almost for every value of $p$ in 
the range $(0,1]$. Lastly, in Figure~\ref{fig:figure_2}f we separate the neurons 
into two groups as follows:
generalist neurons for $i=1,\dots,N/2$, and specialists for $i=N/2+1,\dots, N$. 
We find that: $3/4$ of interneurons are classified as generalists and only $1/4$ as 
specialists; motor neurons are split nearly half and half between generalists and 
specialists; and $2/3$ of sensory and polymodal neurons are specialists while $1/3$ 
of them are generalists (broad functional categories of neurons are compiled 
and provided at~\url{http://www.wormatlas.org/ hermaphrodite/nervous/Neuroframeset.html}, 
Chapter 2.2~\cite{wormatlas}. A classification for every neuron into four broad 
neuron categories follows: (1) interneurons , (2) motor neurons, (3) sensory neurons, 
and (4) polymodal neurons~\cite{wormatlas}).

\begin{figure}[h]
\includegraphics[width=\textwidth]{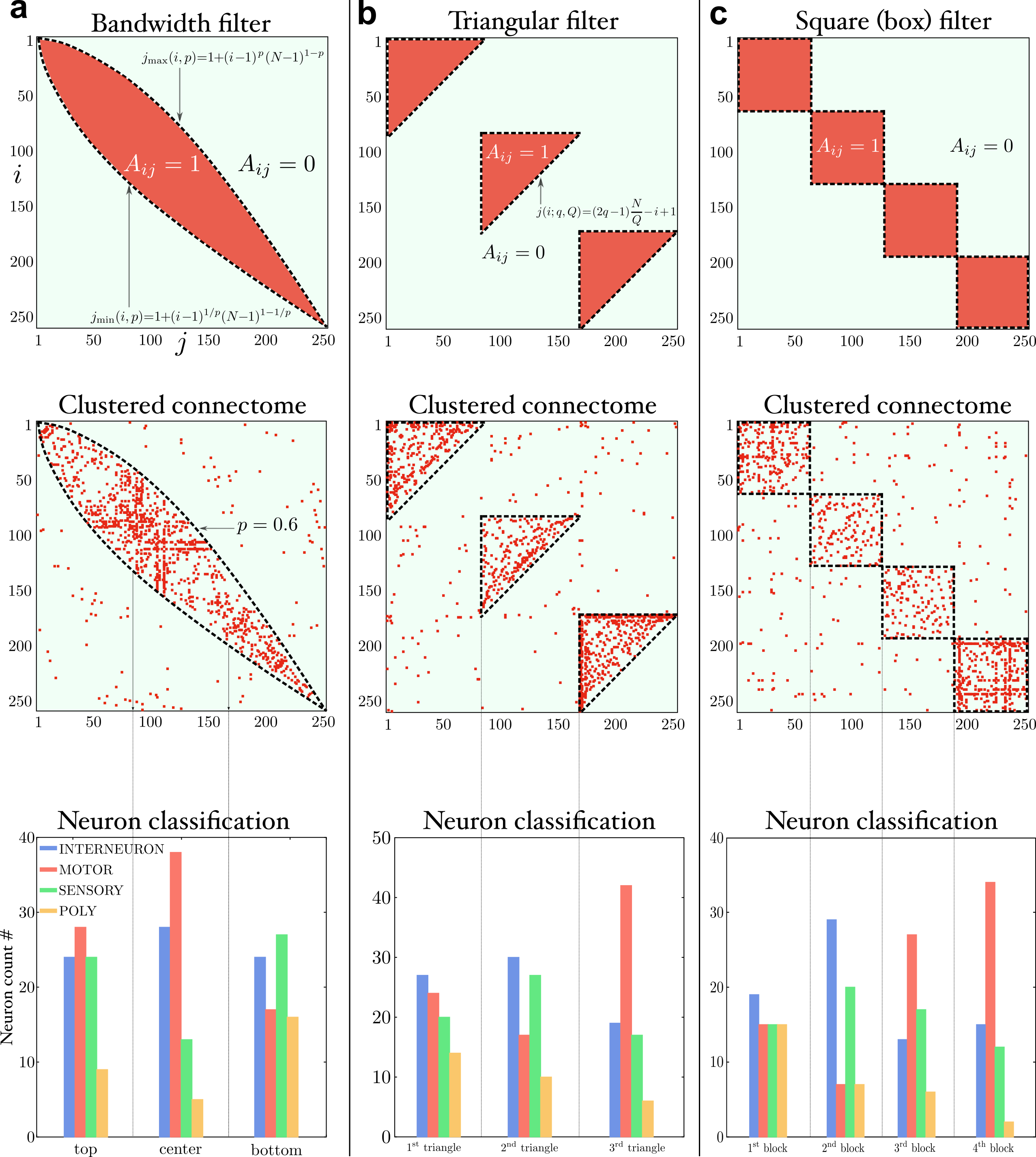} 
\vspace*{-2mm}
\caption{}
\label{fig:figure_3} 
\end{figure}

Last but not least, we present three more types of clustering performed 
on the {\it C.elegans} connectome, by using three more filters: 
the bandwidth filter~\cite{chinn}, shown in Figure~\ref{fig:figure_3}a 
the triangular filter and the square (or box) filter~\cite{jain}, whose mathematical 
properties are discussed in SI sec.~\ref{sec:filters}. 

\begin{figure}[h]
\includegraphics[width=\textwidth]{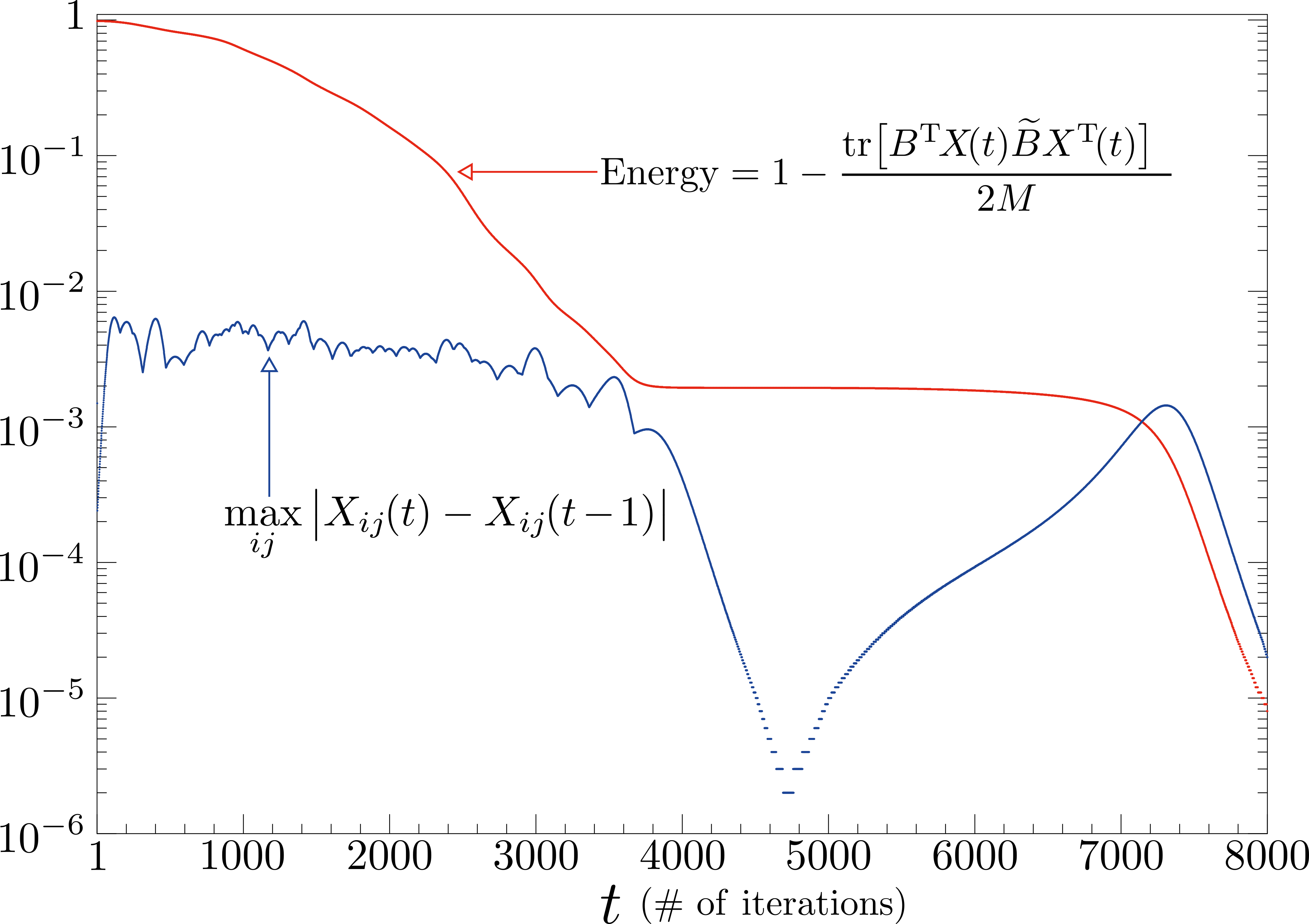} 
\vspace*{-2mm}
\caption{}
\label{fig:figure_4} 
\end{figure}

Furthermore, we notice that if $A$ represents itself the graph of a network, 
then the OPP is equivalent to the graph isomorphism problem~\cite{babai}, 
as exemplified in Figure~\ref{fig:figure_4}, which becomes the graph 
automorphism problem in the special case $A=B$. 
In this latter case, the OPP is equivalent to the problem of mimizing the 
norm of the commutator $E(P)=||[A,P]||^2$. Then, the optimal permutation 
$P_*$ is called a `symmetry of the network' if $E(P_*)=0$, or a 
`pseudosymmetry' if the weaker condition $E(P_*)>0$ holds true~\cite{morone}. 

In conclusion, our analytical and algorithmic results for clustering a matrix 
by solving the optimal permutation problem reveal their importance in that 
their essential features are not contigent on a special form of the matrix 
nor on special assumptions concerning the filters involved. 
We may well expect that it is this part of our work which is most certain 
to find important applications not only in natural science but also in the 
understanding of artificial systems.

\bigskip
\noindent
{\bf \large Data availability}

Data that support the findings of this study are publicly available 
at the Wormatlas database at~\url{https://www.wormatlas.org}.

\noindent
{\bf \large Acknowledgments} 
\noindent

\begin{figure}[h]
\includegraphics[width=\textwidth]{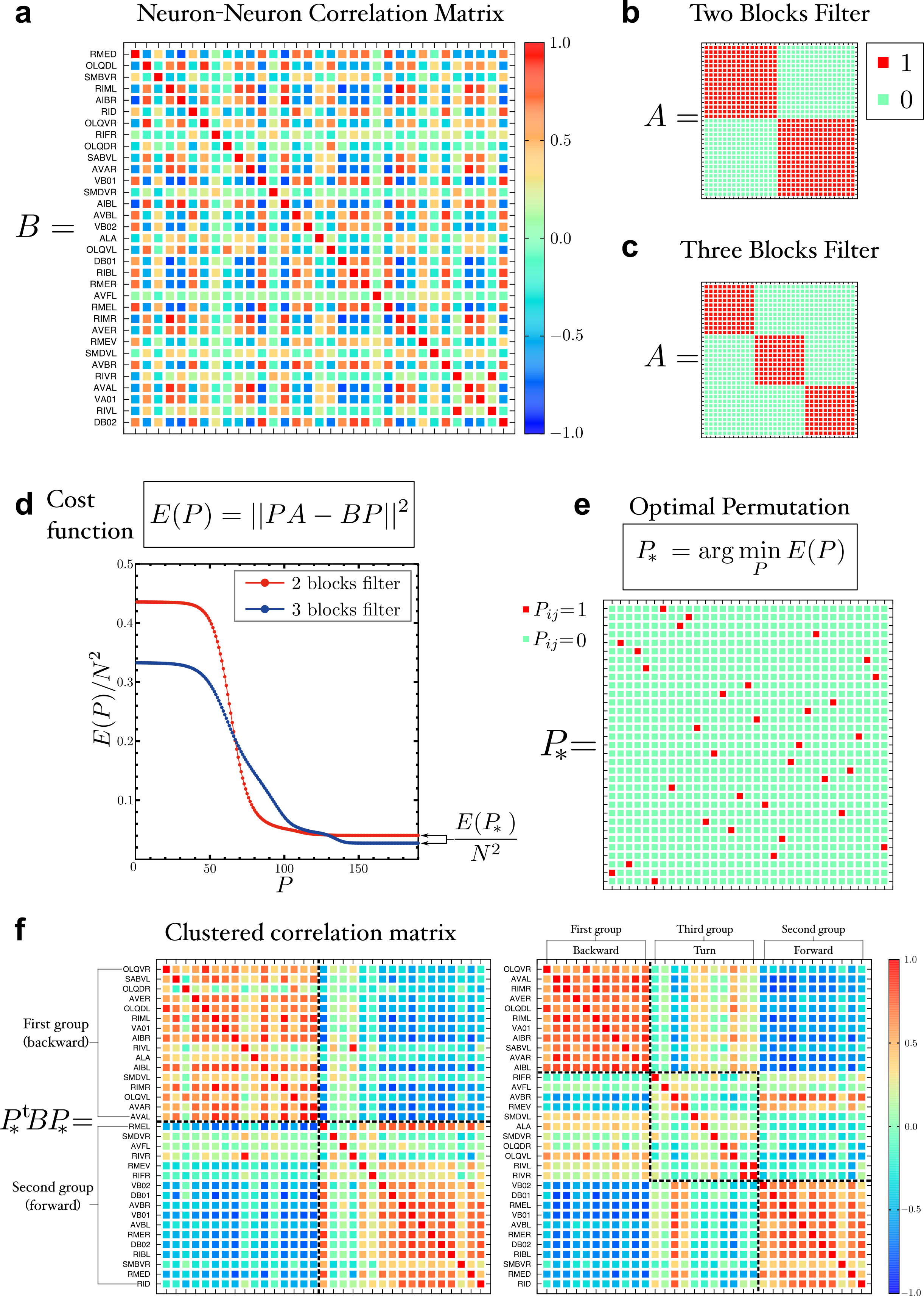} 
\vspace*{-2mm}
\caption{}
\label{fig:figure_1} 
\end{figure}

We thank M. Zimmer for providing the time series used in 
Figure~\ref{fig:figure_1}a.

\noindent
{\bf \large Author contributions}

F. M. designed research, developed the theory, run the algorithm, and wrote the 
manuscript.

\noindent
{\bf \large Additional information}

{\bf Supplementary Information} accompanies this paper. 
%{\bf Supplementary Information} accompanies this paper at \url{https://www.nature.com/}.

\noindent
{\bf \large Competing interests} 

The author declares no competing interests. 

\noindent
{\bf Correspondence and requests for source codes} should be addressed to F. M. 
at: flaviomorone@gmail.com

%%% References

%\clearpage 
%\section{Methods}
%\label{methods}

\clearpage

{\bf Fig.~\ref{fig:figure_1}}.
{\bf Explanation of the optimal permutation problem.} 
{\bf a} Correlation matrix of the neuronal activity of the C. elegans. 
Each entry $C_{ij}$ is the correlation coefficient between the time 
series $x_i^t$ and $x_j^t$ measuring the temporal activities of 
neurons $i$ and $j$ (data are from Ref.~\cite{zimmer}).
{\bf b} Two blocks filter $A$ to be applied to matrix $B$ to perform 
the clustering of $B_{ij}$ into $2$ blocks, each one made up of 
neurons maximally correlated among each other. The two blocks 
are arranged along the main diagonal.
{\bf c} Three blocks filter which, similarly to the filter in {\bf b}, produces 
a clustering of $B$ into $3$ clusters. 
{\bf d} Minimization of the cost function $E(P)$ defined in 
Equation~\eqref{eq:costfunction} for two and three-blocks filters. 
{\bf e} The optimal permutation $P_*$ that solves the OPP defined 
in Equation~\eqref{eq:optimalpermutation} for the correlation matrix 
$B$ shown in {\bf a} and the two-blocks filter $A$ in {\bf b}. 
The permutation $P_*$ is the one that minimizes the cost function 
in {\bf d} (red curve), i.e., $P_*: \min_{P}E(P) = E(P_*)$.
{\bf f} Clustered correlation matrix $B'=P_*^{\rm t}BP_*$ obtained 
by solving the OPP with a two-blocks filter (left panel) and a 
three-blocks filter (right panel).

\bigskip

{\bf Fig.~\ref{fig:figure_2}}.
 {\bf Clustering the {\it C. elegans} connectome through optimal 
permutations.}
{\bf a} The adjacency matrix $B$ of the C. elegans gap-junction 
connectome from ref.~\cite{varshney}. The matrix is binary so its 
entries take two possible values: $B_{ij}=1$ if a gap-junction exists between 
neurons $i$ and $j$, and $B_{ij}=0$ if not.  
{\bf b} The nestedness filter $A$ used to cluster the adjacency matrix $B$ 
defined in {\bf a}. Matrix $A$ is a binary matrix having entries $A_{ij} = 1$ for  
$j <= f(i;p) = N - (i-1)^p(N-1)^{1-p}$ (corresponding to the red 
area extending from the upper left corner to the black dashed line 
defined by the equation $j=f(i;p)$); and $A_{ij}=0$ for $j > f(i;p)$ ( 
corresponding to the complementary light-green area). We choose the 
nestedness exponent $p=0.4$. 
{\bf c} The optimal permutation matrix $P_*$ obtained by solving the OPP 
with the matrices $B$ and $A$ shown in {\bf a} and {\bf b} respectively. 
{\bf d} The clustered adjacency matrix obtained from $B$ by applying 
a similarity transformation with the optimal permutation matrix $P_*$ 
found in {\bf c}, that is $P_*^{\rm t}BP_*$ (left side). 
Right side: clustered adjacency matrices obtained with nine different 
filters having nestedness exponents $0.1\leq p\leq 1.0$.
{\bf e} The packing fraction of {\it C.elegans} connectome $\phi(p)$ (red dots), 
defined by Equation~\eqref{eq:packingfraction}, as a function of the nestedness 
exponent $p$, as compared to the average packing fraction $\phi_{\rm ran}(p)$ 
(black crosses) of a randomized connectome with the same degree sequence 
(error bars are s.e.m. over $10$ realizations of the configurational model). 
The inset shows the difference $\phi(p) - \phi_{\rm ran}(p)$ as a function 
of $p$, that has a maximum equal to $\sim 0.1$ for $p=0.5$. 
{\bf f} Classification of neurons into generalists and specialists as explained 
in the main text.

 \bigskip

{\bf Fig.~\ref{fig:figure_3}}. {\bf Various clustering types}.
{\bf a} The bandwidth filter (upper panel) and the clustered 
 {\it C.elegans}  connectome (middle pannel). The classification 
 of neurons in the band (bottom panel) shows that motor neurons 
 are predominantly located in the central part of the band, which 
 is the part with the largest bandwidth; sensory neurons instead 
 are located mostly in the extremal parts of the band (upper and lower ends); 
 while interneurons are almost evenly distributed along the band; 
 polymodal neurons are mostly residing in the bottom part of the band. 
 {\bf b} Triangular filter (upper panel) and the corresponding 
 triangular clustering of the connectome into $3$ triangular blocks 
 (middle panel). The visually most prominent feature in the neuron 
 classification (lower panel) is that half of the motor neurons tend 
 to cluster all into one trianglular block. 
 {\bf c} Square (or box) filter (upper panel) and the corresponding 
 block-diagonal clustering of the connectome into $4$ square 
 blocks (middle panel). The neuron classification in the lower panel 
 shows a visibible segregation of interneurons populating mostly 
 the $1^{\rm st}$ and $2^{\rm nd}$ blocks from motor neurons 
 situated mostly in the $3^{\rm rd}$  and $4^{\rm th}$ blocks. 
 
 \bigskip
 
{\bf Fig.~\ref{fig:figure_4}}.
{\bf Solution to the graph isomorphism problem.} 
We solve the graph isomorphism problem by using the adjacency 
matrix $B$ of the {\it C. elegans} connectome, depicted in 
Figure~\ref{fig:figure_2}a, and a matrix $\widetilde{B}$ obtained 
from $B$ by a similarity transformation $\widetilde{B}=P_{\rm r}BP_{\rm r}^{\rm t}$, 
where $P_{\rm r}$ is a random permutation matrix with no fixed 
points (also called a derangement). In other words, $\widetilde{B}$ is 
permutation similar to $B$, hence the two graphs represented by 
$B$ and $\widetilde{B}$ are isomorphic by construction. 
Of course, this information is not exploited by our algorithm, which 
has, {\it a priori}, no clue on how the matrix $\widetilde{B}$ has been 
generated and wether an isomorphism exists between $B$ and $\widetilde{B}$ at all. 
The fact that the minimum of the energy function (red dots) goes to zero 
as the number of iterations $t$ of the algorithm increases means that our algorithm is able to 
determine that graphs $B$ and $\widetilde{B}$ are indeed isomorphic. 
Moreover, the optimal permutation $P_*$ returned by the algorithm is an 
explicit example of graph isomorphism. We note that $P_*$ and $P_{\rm r}$ 
need not to be necessarily the same permutation matrix. This is due to the 
existence of symmetries (i.e. automorphisms) of the matrix $B$. 
These symmetries are permutation matrices $S$ that commute with $B$, 
so that we can write $B=SBS^{\rm t}$ for any $S$ such that $[B,S]=0$. 
%meaning that the adjacency matrix $B$ is invariant under a permutation $S$ 
%of its individual nodes.  
Therefore, we will have as many solutions $P_*$ to the graph isomorphism 
problem as simmetries $S$ there are in the connectome $B$. 
Thus, we can retrieve the original permutation $P_{\rm r}$ only up to an 
automorphism of $B$, i.e, $P_* = P_{\rm r}S$. 
The blue dots represent the maximum difference between the entries 
$X_{ij}(t)$ at iteration $t$ and $X_{ij}(t-1)$ at the previous step $t-1$.

\clearpage
%%%%%%%%%%%%%%%
%FIGURES
%%%%%%%%%%%%%%%

%Correlation clustering

%Network clustering

%More clustering

%Graph isomorphism 

\clearpage

\renewcommand{\figurename}{Supplementary Figure}
\renewcommand{\tablename}{Supplementary Table}

\centerline{ \bf \large Supplementary Information for:}

\centerline{\bf Clustering matrices through optimal permutations}

\medskip

\centerline{Flaviano Morone}

\tableofcontents

\clearpage

\section{The Optimal Permutation Problem (OPP)}
\label{sec:theory}
We consider two square matrices $A,B\in \mathcal{M}_{N\times N}(\mathbb{R})$, 
where $\mathcal{M}_{N\times N}(\mathbb{R})$ is the vector space of $N\times N$ 
real matrices, and the vector space $\mathcal{P}_{N\times N}$ of $N\times N$ 
permutation matrices $P$ having their entries $P_{ij}\in\{0,1\}$ that satisfy 
the constraints of row and column sums equal to one: $\sum_i P_{ij}=\sum_jP_{ij}=1$. 
Next we define $\Delta(P): \mathcal{P}_{N\times N}\to \mathcal{M}_{N\times N}$ 
to be the function
\be
\Delta(P) = PA-BP\ .
\ee
We call $A$ the `filter' (or `template') matrix, and $B$ the `input' matrix. 
We define the inner product $\langle L|R\rangle$ between two matrices 
$L,R\in\mathcal{M}_{N\times N}$ by means of the following formula:
\be 
\langle L|R\rangle = {\rm tr}\left(R^{\rm T}L\right)\ ,
\ee
where ${\rm tr}$ indicates the trace operation: ${\rm tr}(A)=\sum_{i=1}^NA_{ii}$.
Then, the norm of $\Delta(P)$ can be computed as:
\be 
\begin{aligned}
||\Delta(P)||^2 &= \langle \Delta(P)|\Delta(P)\rangle = 
{\rm Tr}\left(A^{\rm T}P^{\rm T}PA - A^{\rm T}P^{\rm T}BP 
- P^{\rm T}B^{\rm T}PA + P^{\rm T}B^{\rm T}BP\right) =\\
&= ||A||^2 + ||B||^2 - 2{\rm Tr}\left(B^{\rm T}PAP^{\rm T}\right) 
= ||A||^2 + ||B||^2 - 2{\rm Tr}\left[\left(BPA^{\rm T}\right)^{\rm T}P\right]=\\
&= ||A||^2 + ||B||^2 - 2\langle P| Q(P)\rangle\ ,
\end{aligned}
\label{eq:norm}
\ee
where we have introduced the `overlap' matrix $Q(P)$, which is defined 
as follows:
\be
Q(P) = BPA^{\rm T}\ .
\ee
The quantity we want to optimize over is precisely the inner product 
$\langle P| Q(P)\rangle$. Therefore, we define the objective (or energy) 
function $E(P)$ of our problem as 
\be
E(P) = -\frac{1}{2}\langle P| Q(P)\rangle\ ,
\label{eq:energy}
\ee
where the factor $1/2$ has been chosen for future convenience. 
The optimal permutation problem (OPP) is defined as the problem 
of finding the global minimum (or ground state) of the energy 
function in Equation~\eqref{eq:energy}:
\be
\boxed{\ 
\begin{aligned}
{\rm OPP}\ \vcentcolon  =\ &{\rm minimize}\  E(P)\\
&{\rm s.\ t.}\ P\in \mathcal{P}\ .
\end{aligned}
\ }
\label{eq:opp}
\ee

In order to determine the solution to the OPP given by~\eqref{eq:opp} 
we take a statistical physics approach by introducing a fundamental 
quantity called partition function $Z(\beta)$, defined by the following 
summation: 
\be
Z(\beta) = \sum_{P\in \mathcal{P}} e^{-\beta E(P)}\ ,
\label{eq:partitionfunction2}
\ee
where the notation $\sum_{P\in \mathcal{P}}$ indicates the sum 
over all $N\times N$ permutation matrices $P$. 
The parameter $\beta$ in Equation~\eqref{eq:partitionfunction2} 
represents, in the statistical physics interpretation of the problem, 
the inverse of the `temperature' of the system. 
We notice that the sum in Equation~\eqref{eq:partitionfunction2} involves 
$N!$ terms, and thus grows as the factorial of the system size, $Z\sim O(N!)$, 
rather than displaying the peculiar exponential growth, $Z\sim O(e^N)$, 
that appears in the study of the thermodynamic limit of many-body classic 
and quantum systems. 

The global minimum of the objective function in Equation~\eqref{eq:energy} 
corresponds, physically, to the `ground-state' of the system. But a physical 
system exists in its ground state only at zero temperature (by the third law 
of thermodynamics), and thus the exact solution to our optimization 
(i.e. minimization) problem can be computed by taking the zero temperature 
limit (which is mathematically tantamount to send $\beta\to\infty$) of the 
partitition function defined by Equation~\eqref{eq:partitionfunction2}. 
Specifically, the minimum of $E(P)$ is given by:
\be
\min_{P\in \mathcal{P}}E(P) = E(P_*) = 
\lim_{\beta\to\infty}-\frac{1}{\beta}\log Z(\beta)\ . 
\ee

Since the partition function in Equation~\eqref{eq:partitionfunction2} can be easily 
calculated when all $P_{ij}$ appear linearly in the argument of the 
the exponential, a good idea is to write the quadratic term which in the energy 
connects two variables $P_{ij}$ and $P_{k\ell}$ on different links $i\to j$ and 
$k\to \ell$ as an integral over disconnected terms. In order to achieve this 
result, we insert the $\delta$-function
\be
\delta(X_{ij}-P_{ij}) = \frac{1}{2\pi \imath}\int d J_{ij}\ {\rm e}^{J_{ij}(P_{ij}-X_{ij})}\ ,
\ee
where the integration over $J_{ij}$ runs along the imaginary axis, into the 
representation of the partition function~\cite{zinn}:
\be
Z(\beta) \propto  \sum_{P\in \mathcal{P}}
\int\prod_{ij}d X_{ij}\int \prod_{ij} J_{ij}\ {\rm e}^{-\beta E(X) 
\ +\ \sum_{ij}J_{ij}(P_{ij}-X_{ij})}\ ,
\ee
To proceed further in the calculation, we enforce the costraint on the 
column sums: 
\be 
\sum_{i}P_{ij}=\sum_{i}X_{ij}=1\ ,\ \ \  \forall j\ ,
\ee
by inserting $N$ $\delta$-functions
\be 
\delta\Big(\sum_{i}X_{ij}-1\Big) =
\frac{1}{2\pi \imath}\int d z_j\ {\rm e}^{-z_j\left(\sum_iX_{ij}-1\right)}\ ,\ \ \ j=1,\dots,N\ ,
\ee
into the partition function:
\be
Z(\beta) \propto \sum_{P\in \mathcal{P'}}
\int\prod_{ij}d X_{ij}\int \prod_{ij} J_{ij}\ \int\prod_j z_j\ 
{\rm e}^{-\beta E(X) \ +\ \sum_{ij}J_{ij}(P_{ij}-X_{ij})\ -\ \sum_jz_j\left(\sum_iX_{ij}-1\right)}\ ,
\label{eq:partitionfunction3}
\ee
where $\mathcal{P'}$ indicates the vector space of $N\times N$ right stochastic matrices
$P$ with integer entries $P_{ij}\in\{0,1\}$, that is, matrices with each row summing to one: 
$\sum_{j=1}^NP_{ij}=1$ (but no costraint on the column sums). 
Then, summation over the variables $P_{ij}$ is straightforward, and we find:
\be
\sum_{P\in \mathcal{P'}}{\rm e}^{\sum_{ij}J_{ij}P_{ij}} = \prod_i\sum_j {\rm e}^{J_{ij}}\ .
\ee
Introducing the function $F(X,J,z)$ defined by:
\be
F(X,J,z;\beta) = E(X) + \frac{1}{\beta} \langle X|J\rangle + 
\frac{1}{\beta}\sum_jz_j\Big(\sum_iX_{ij}-1\Big) - 
\frac{1}{\beta}\sum_i\log\sum_j {\rm e}^{J_{ij}},  
\ee
we can write the partition function Equation~\eqref{eq:partitionfunction3} as
\be
Z(\beta) \propto\ \int \prod_{ij}d X_{ij}d J_{ij}\prod_jdz_j\ {\rm e}^{-\beta F(X,J,z;\beta)}\ ,
\ee
which can be evaluated by the steepest descent method in the limit of zero 
temperature (i.e. $\beta\to\infty$). The saddle point equations are obtained 
by differentianting $F$ with respect to $X_{ij}$, $J_{ij}$, and $z_j$:
\be 
\begin{aligned}
\frac{\partial F}{\partial X_{ij}} &= -\frac{1}{2}\frac{\partial}{\partial X_{ij}}
\Big[\sum_{k\ell}X_{k\ell}Q_{k\ell}(X)\Big] + \frac{1}{\beta}J_{ij}+
\frac{1}{\beta}z_j = 0\ ,\\
\frac{\partial F}{\partial J_{ij}} &= \frac{1}{\beta}X_{ij} - 
\frac{1}{\beta}\sum_k\frac{1}{\sum_{\ell}{\rm e}^{J_{k\ell}}}\sum_{\ell}
\frac{\partial}{\partial J_{ij}}{\rm e}^{J_{k\ell}} = 0\ ,\\
\frac{\partial F}{\partial z_j} & = \frac{1}{\beta}\Big(\sum_iX_{ij}-1\Big) = 0\ .
\end{aligned}
\label{eq:saddlepoint2}
\ee
The solution to the saddle point Equations~\eqref{eq:saddlepoint2} is given by:
\be 
\begin{aligned}
J_{ij} &= \frac{\beta}{2}\left(BXA^{\rm T} + B^{\rm T}XA\right)_{ij} - z_j\ ,\\
X_{ij} &= \frac{{\rm e}^{J_{ij}}}{\sum_k {\rm e}^{J_{ik}}}\ ,\\
1 &= \sum_iX_{ij}\ .
\end{aligned}
\label{eq:solutionSP}
\ee
Notice that the solution $X_{ij}$ satisfies automatically the condition 
of having columns summing to one: $\sum_j X_{ij}=1$, $\forall i$, as it 
should. Opposed to this, are the constraints on the row sums, 
$\sum_i X_{ij}=1$ $\forall j$, which are taken into account by the 
Lagrange multipliers $z_j$. 

Next, we eliminate $J_{ij}$ in favor of $X_{ij}$ and we make the 
constraints on the row and column normalizations manifest in the 
final solution. Introducing the matrix $W(X)$ defined by
\be
W_{ij}(X) =  \frac{1}{2}\left(BXA^{\rm T}+B^{\rm T}XA\right)_{ij}\ ,
\ee
we can write $J_{ij}$ as $J_{ij}=\beta W_{ij}(X) -  z_j$. Thus, $X_{ij}$ 
in Equation~\eqref{eq:solutionSP} takes the form
\be
X_{ij} = \frac{{\rm e}^{\beta W_{ij}(X) -  z_j}}{\sum_k {\rm e}^{ \beta W_{ik}(X) -  z_k}}\ .
\label{eq:solutionX}
\ee
We notice that Equation~\eqref{eq:solutionX} is invariant under global 
translations of the form
\be
z_j\ \to\ z_j + \zeta\ ,\ \ \ \forall j\ ,
\ee
for arbitary values of $\zeta$. This symmetry is not unexpected and can be traced 
back to the fact that out of the $2N$ constraints on the row and columns normalization, 
only $2N-1$ of them are linarly independent, since the sum of all entries must be equal to 
$N$, i.e., $\sum_{ij}X_{ij}=N$.  
This translational symmetry can be eliminated, for example, by choosing $\zeta$ 
in such a way that:
\be
\sum_iz_i = 0\ . 
\ee
In general, the Lagrange multipliers $z_j$ in Equation~\eqref{eq:solutionX} can 
be eliminated, in principle, by imposing the constraints $\sum_i X_{ij}=1$, i.e., by 
solving th following system of equations:
\be
z_j = \log\sum_i \left(\frac{ {\rm e}^{\beta W_{ij}}}{\sum_k e^{\beta W_{ik}-z_k}}\right)\ ,\ \ \ 
j =1,\dots, N\ .
\ee
We define, just for future notational convenience, the variable  $Y_{ij}(X)$ 
as follows
\be 
Y_{ij}(X) = {\rm e}^{\beta W_{ij}(X)}\ .
\label{eq:expw}
\ee
Then, we can make the normalization constraints manifest by defining two 
vectors: a right vector with components $V_j$
\be
V_j = {\rm e}^{-z_j}\ , 
\ee
and a left vector with components $U_i$
\be 
U_i = \frac{1}{\sum_j Y_{ij}(X)\ V_j}\ ,
\label{eq:leftvector}
\ee
whereby we can rewrite Equation~\eqref{eq:solutionX} as 
\be
\boxed{\ X_{ij}\ =\  U_i(X)\ Y_{ij}(X) V_j(X)\ }\ , 
\label{eq:main}
\ee
where $V_j(X)$ can be calculated consistently with $U_i(X)$ using the 
following equations:
\be 
V_j = \frac{1}{\sum_i U_i\ Y_{ij}(X)}\ .
\label{eq:rightvector}
\ee
We notice that Equations~\eqref{eq:leftvector} and~\eqref{eq:rightvector} 
are nothing but the Sinkhorn-Knopp equations~\cite{sinkhorn1, sinkhorn2} 
to rescale all rows and all columns of a matrix with strictly positive entries 
(as is indeed each term $Y_{ij}(X) = {\rm e}^{\beta W_{ij}(X)}>0$ in the the 
present case) to sum to one.

\section{Algorithm to solve the saddle point equations and find the optimal permutation}
\label{sec:algorithm}
In order to find the matrix $X_*$ that solves equations~\eqref{eq:saddlepoint2} 
we set up an algorithm defined by the following iterative procedure. 
First of all, we need to introduce a regularized kernel $Y(X; \epsilon)$ as follows 
\be
\begin{aligned}
W_{ij}(X; \epsilon) &= W_{ij}(X)+\beta\epsilon X_{ij}\ ,\\
Y_{ij}(X; \epsilon) &= Y_{ij}(X){\rm e}^{\beta\epsilon X_{ij}}\ ,
\end{aligned}
\label{eq:regularization}
\ee
with $\epsilon>0$ being a smoothing parameter to be send eventually to zero. 
In all our experiments we set $\epsilon= 10$ at the start, and then decrease it
by one, $\epsilon \to \epsilon -1$ until $\epsilon = 0$, after each completion of 
the following routine.
\begin{itemize}
\item {\bf 1)} Initialize $X_{ij}^{(t=0)}\equiv X_{ij}^{(0)}$ at time zero to a uniform 
matrix as: $X_{ij}^{(0)}=1/N$. 
\item {\bf 2)} Calculate the quantity: 
\be
Y_{ij}\big(X^{(0)}; \epsilon; \alpha\big) \equiv Y_{ij}^{(0)} = 
\exp\left[\alpha\beta W_{ij}\big(X^{(0)};\epsilon\big) + (1-\alpha)\log\big(X^{(0)}\big)\right]\ .
\ee
We choose $\alpha = 10^{-3}$ and $\beta = 10$ (the parameter $\alpha$ is a 
`dumping' factor which helps the convergence of the algorithm).
%and preventing overflowing values to occur). 
%
\item {\bf 3)} Calculate $U_i$ and $V_j$ as follows: 
\begin{itemize}
\item {\bf a)} Initialize $U_i^{(0)}=1$ $\forall\ i$.
\item {\bf b)} Compute $V_j^{(1)}$ and $U_i^{(1)}$ using the following equations:
\be
\begin{aligned}
V_j^{(1)} &= \frac{1}{\sum_i U_i^{(0)} Y_{ij}^{(0)}}\ ,\\
U_i^{(1)} &= \frac{1}{\sum_j Y_{ij}^{(0)} V_j^{(1)}}\ .
\end{aligned}
\ee
\item {\bf c)} Calculate the quantity $\delta$ defined as follows:
\be
\delta = \max_i \left| U_i^{(1)} - U_i^{(0)}\right|\ .
\ee
\item {\bf d)} If $\delta > 10^{-5}$ then start over from step {\bf b}); 
otherwise {\bf return} $U_i$ and $V_j$. 
\end{itemize}
\item {\bf 4)} Update $X$ by computing $X_{ij}^{(1)}$ as
\be 
X_{ij}^{(1)} = U_i Y_{ij}^{(0)} V_j\ .
\ee
\item {\bf 5)} Calculate the quantity $\Delta$ defined by 
 \be
\Delta = \max_{ij} \left|X_{ij}^{(1)} - X_{ij}^{(0)}\right|\ .
\ee
\item {\bf 6)} If $\Delta > 10^{-5}$ then start over from step {\bf 2)}; 
otherwise {\bf return} $X_{ij}$. 
\end{itemize}

The output matrix $(X_*)_{ij}$ is not a permutation matrix, 
but only a double-stochastic matrix, that is a matrix whose entries 
are real numbers $X_{ij}\in[0,1]$ that satisfies the double normalization 
condition on row and column sums: $\sum_iX_{ij}=\sum_jX_{ij}=1$. 
To find the solution of the OPP we should take the zero temperature 
limit by sending $\beta\to\infty$: in this limit the solution matrix 
$X_*$ is projected onto one of the $N!$ vertices of the Birkhoff 
polytope, which represents the optimal permutation matrix $P_*$ 
that solves the OPP. 
In order to find $P_*$ numerically, we use a simple method which 
consists in finding a solution $X_*$ at large, but finite, $\beta$ 
(we use $\beta=10$), followed by a hard thresholding of the matrix 
entries $(X_*)_{ij}$ defined by:
\be
(P_*)_{ij} = 
\left\{
\begin{matrix}
1 & {\rm if\ } (X_*)_{ij}\geq 0.99\\
0 & {\rm otherwise}
\end{matrix}
\right.
\ .
\ee

\section{Filter matrices}
\label{sec:filters}

Having discussed how to implement the algorithm, we present, 
next, several types of filter matrices $A$ that we used in our 
clustering experiments.

\subsection{Nestedness filter}
\begin{figure}[h!]
\includegraphics[width=0.9\textwidth]{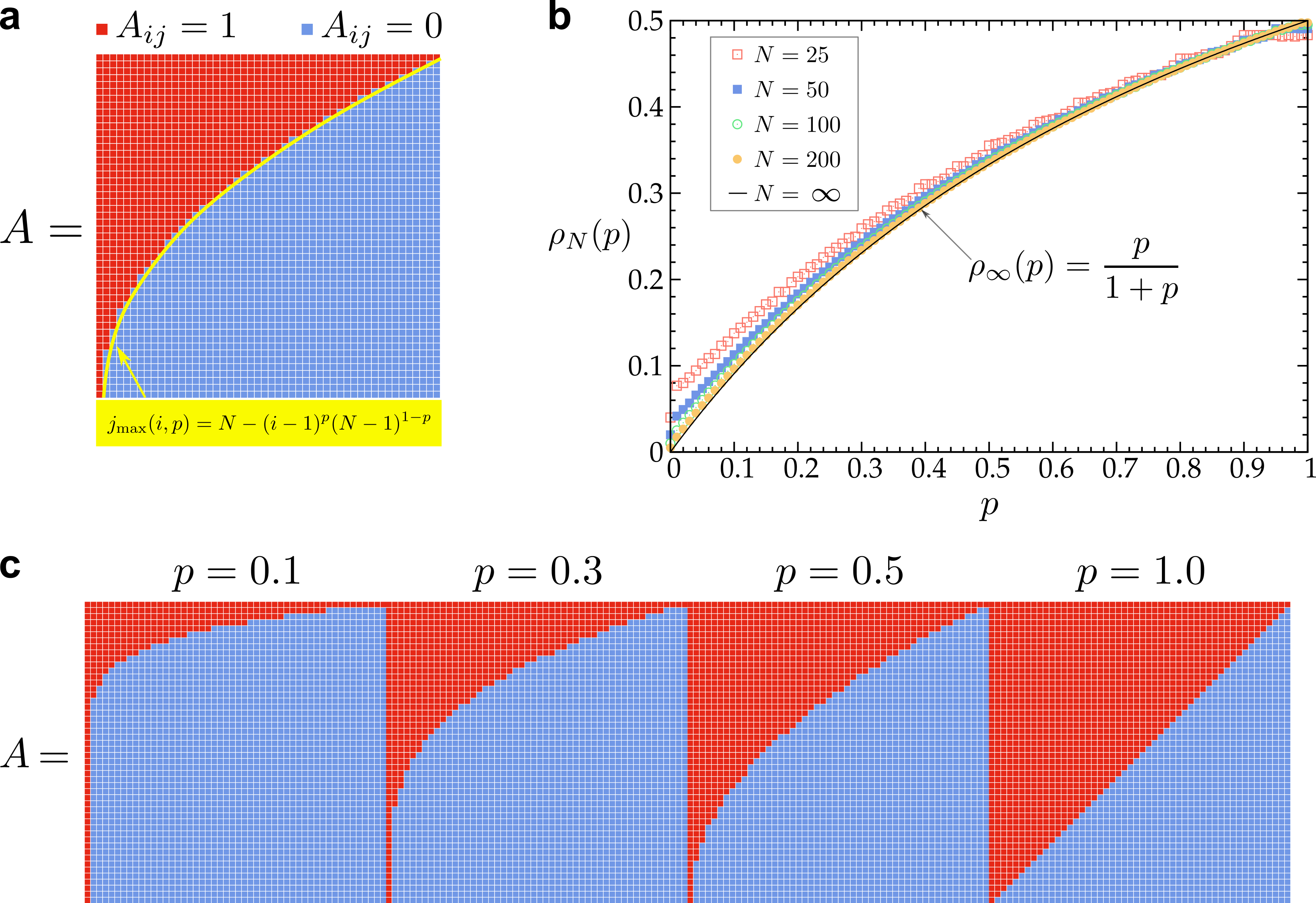}
\centering
\caption{{\bf Nestedness filter}. }
\label{fig:nest}
\end{figure}

The nestedness filter is described by a matrix $A$ whose 
nonzero entries $A_{ij}$ are equal to $1$ when the following 
condition is satisfied:
\be
A_{ij} = 1\ \ \ {\rm for}\ \ \ 1\leq i \leq N\ \ \ {\rm and}\ \ j\in[1,j_{\rm max}(i,p)]\ ,
\label{eq:nfilter}
\ee
where $j_{\rm max}(i,p)$ is given by:
\be
j_{\rm max}(i,p) = N -(i-1)^p(N-1)^{1-p}\ ,
\ee
as shown in Fig.~\ref{fig:nest}a. 
The parameter $p\in[0,1]$ quantifies the nestedness of the matrix $A$.
Specifically, low values of $p$ correspond to a matrix $A$ with a highly 
nested structure. Opposite to this, large values of $p$, i.e. 
$p\sim 1$, describe profiles of low nestedness, as depicted in 
Fig.~\ref{fig:nest}a,c 

The density $\rho(p)$ of the filter matrix $A$ defined by Equation~\eqref{eq:nfilter} 
is given by 
\be
\rho(p) = \frac{1}{N^2}\sum_{i=1}^N\sum_{j=1}^{j_{\rm max}(i,p)}1\ ,
\ee
which, in the limit $N\to\infty$, becomes:
\be
\boxed{\ \rho(p) = \frac{p}{1+p}\ }\ .
\label{eq:densitynested}
\ee
The finite $N$ behavior of $\rho(p)$ together with its $N\to\infty$ limit 
is showm in Fig.~\ref{fig:nest}b.

\subsection{Band filter}
\begin{figure}[h!]
\includegraphics[width=0.9\textwidth]{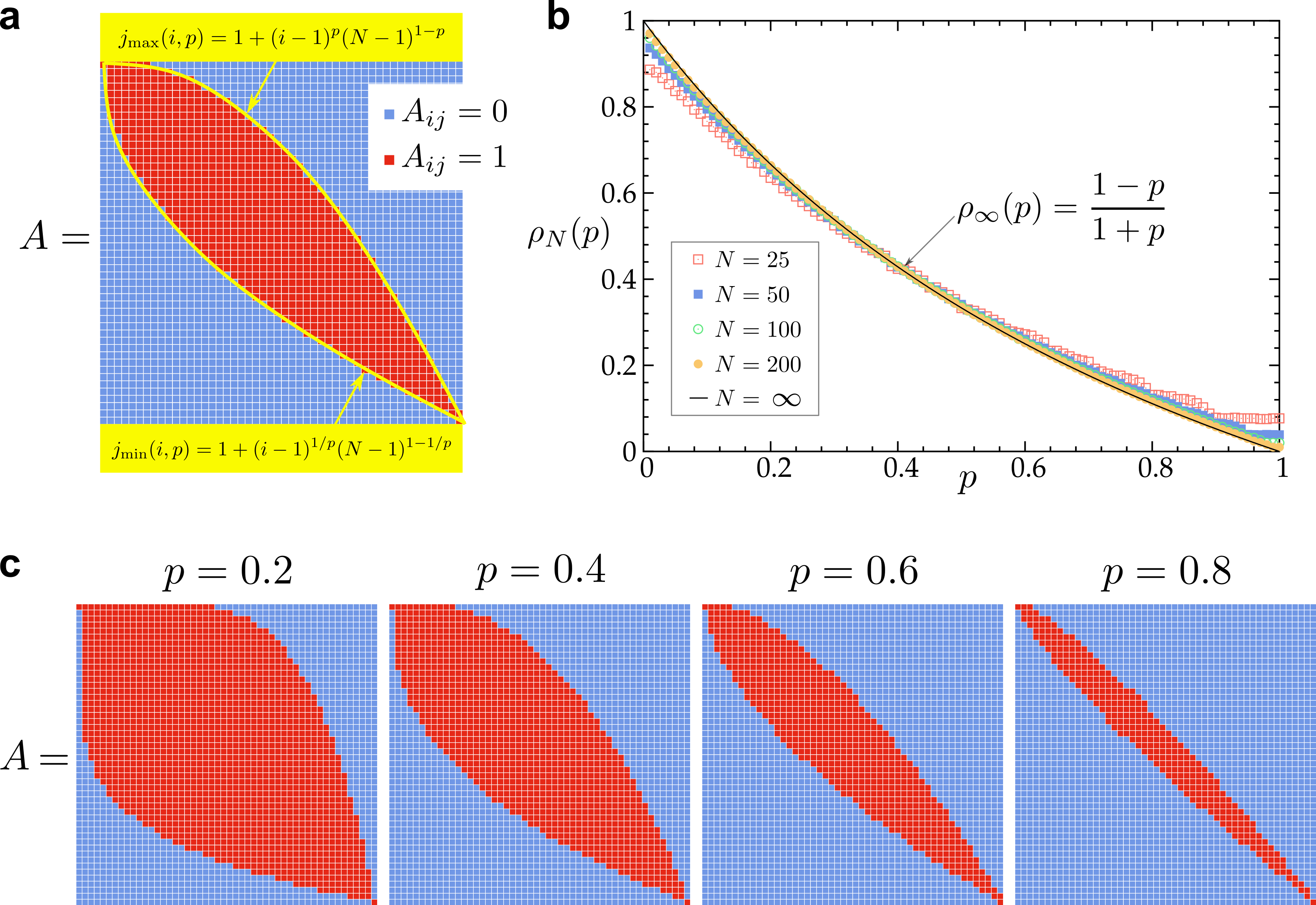}
\centering
\caption{{\bf Band filter}. }
\label{fig:band}
\end{figure}

The band filter is a matrix $A$ whose entries $A_{ij}\in\{0,1\}$ 
are defined by
\be
A_{ij} = \left\{
\begin{matrix}
1 & 
\begin{matrix}
{\rm for}\ 1\leq i \leq N \\
j_{\rm min}(i, p)\leq j\leq j_{\rm max}(i, p)
\end{matrix}
   &\\
0 & {\rm otherwise}
\end{matrix}
\right.\ ,
\label{eq:bfilter}
\ee
where 
\be
\begin{aligned}
j_{\rm min}(i, p) &= 1 + (i-1)^{1/p}(N-1)^{1-1/p}\ ,\\
j_{\rm max}(i, p) &= 1 + (i-1)^{p}(N-1)^{1-p}\ , 
\end{aligned}
\ee
where $p$ is a parameter that controls the width of the band, hence 
we call $p$ the {\bf bandwidth} exponent. 
The band filter in Equation~\eqref{eq:bfilter} has nonzero entries 
comprised in a band delimited by $j_{\rm min}(i, p)$ and 
$j_{\rm max}(i, p)$ for $i=1,\dots, N$. 
The density $\rho(p)$ of $A$ is defined as the fraction of entries 
contained inside the band:
\be
\rho(p) = \frac{1}{N^2}\sum_{i=1}^N\sum_{j=j_{\rm min}(i,p)}^{j_{\rm max}(i,p)}1\ .
\ee
For $N\to \infty$, the density $\rho(p)$ evaluates
\be
\boxed{\ 
\rho(p) =  \frac{1-p}{1+p}\ }\ .
\label{eq:banddensity}
\ee
The finite $N$ behavior of $\rho(p)$ along with the $N\to\infty$ 
limit given by Equation~\eqref{eq:banddensity} are showm in 
Figure~\ref{fig:band}b.

\begin{figure}[h!]
\includegraphics[width=0.65\textwidth]{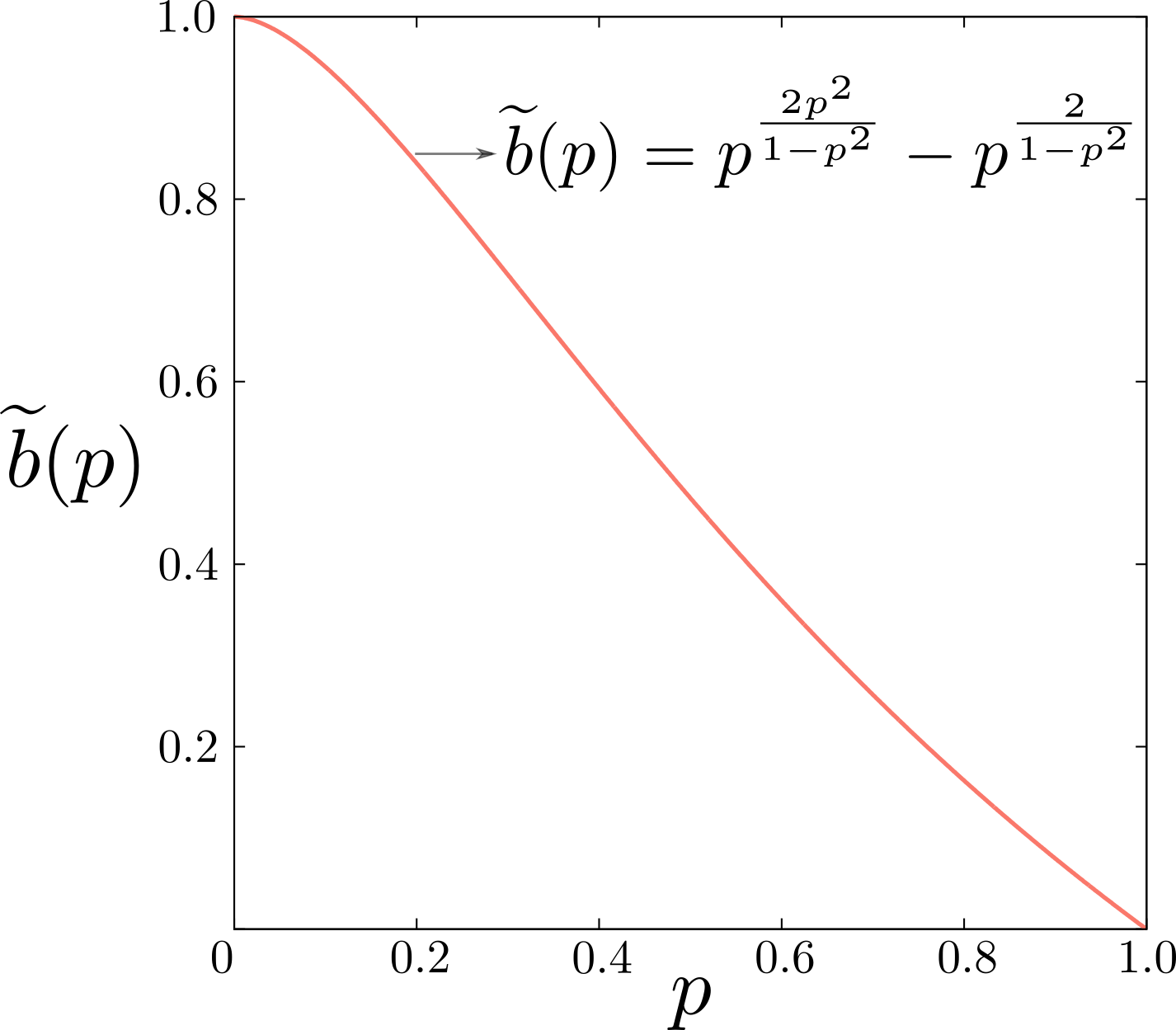}
\centering
\caption{{\bf Rescaled bandwidth}. }
\label{fig:rescaledbandwidth}
\end{figure}

A useful quantity to characterize the shape of the band filter is the 
{\bf bandwidth} $b(p)$, which is defined by
\be
b(p) = \max_{i}[j_{\rm max}(i,p) - j_{\rm min}(i,p)]\ .
\ee
Let us define the rescaled coordinate $x$ taking values 
in the range $x\in[0,1]$ as:
\be
x = \frac{i-1}{N-1}\ ,
\ee
whereby we can write the difference $j_{\rm max}-j_{\rm min}$ as
\be
j_{\rm max}(i,p) - j_{\rm min}(i,p) = (N-1)(x^p - x^{1/p})\ . 
\ee
Next we define the rescaled bandwidth $\widetilde{b}(p)$ as 
\be
\widetilde{b}(p) = \frac{b(p)}{N-1}\ .
\ee
Thus, in the large $N$ limit we can approximate $x$ to a continuous 
variable and thus estimate $\widetilde{b}(p)$ as follows:
\be
\widetilde{b}(p) = x_*(p)^p - x_*(p)^{1/p}\ ,
\label{eq:btilde}
\ee
where $x_*(p)$ is the solution to the following equation:
\be
\frac{d}{dx}(x^p - x^{1/p})\big|_{x=x_*(p)}=0\ , 
\ee
that is:
\be
x_*(p) = p^{\frac{2p}{1-p^2}}\ .
\label{eq:xp}
\ee
Substituting Equation~\eqref{eq:xp} into Equation~\eqref{eq:btilde} 
we obtain the explicit form of the rescaled bandwidth as a function 
of $p$
\be
\boxed{\ 
\widetilde{b}(p) = p^{\frac{2p^2}{1-p^2}} - p^{\frac{2}{1-p^2}}\ }\ ,
\ee
which is shown in Fig.~\ref{fig:rescaledbandwidth}
%%%%%%%%%%%%%%%%%%%%
%%%%%%%%%%%%%%%%%%%%

\subsection{Square filter and Triangle filter}
\begin{figure}[h!]
\includegraphics[width=0.9\textwidth]{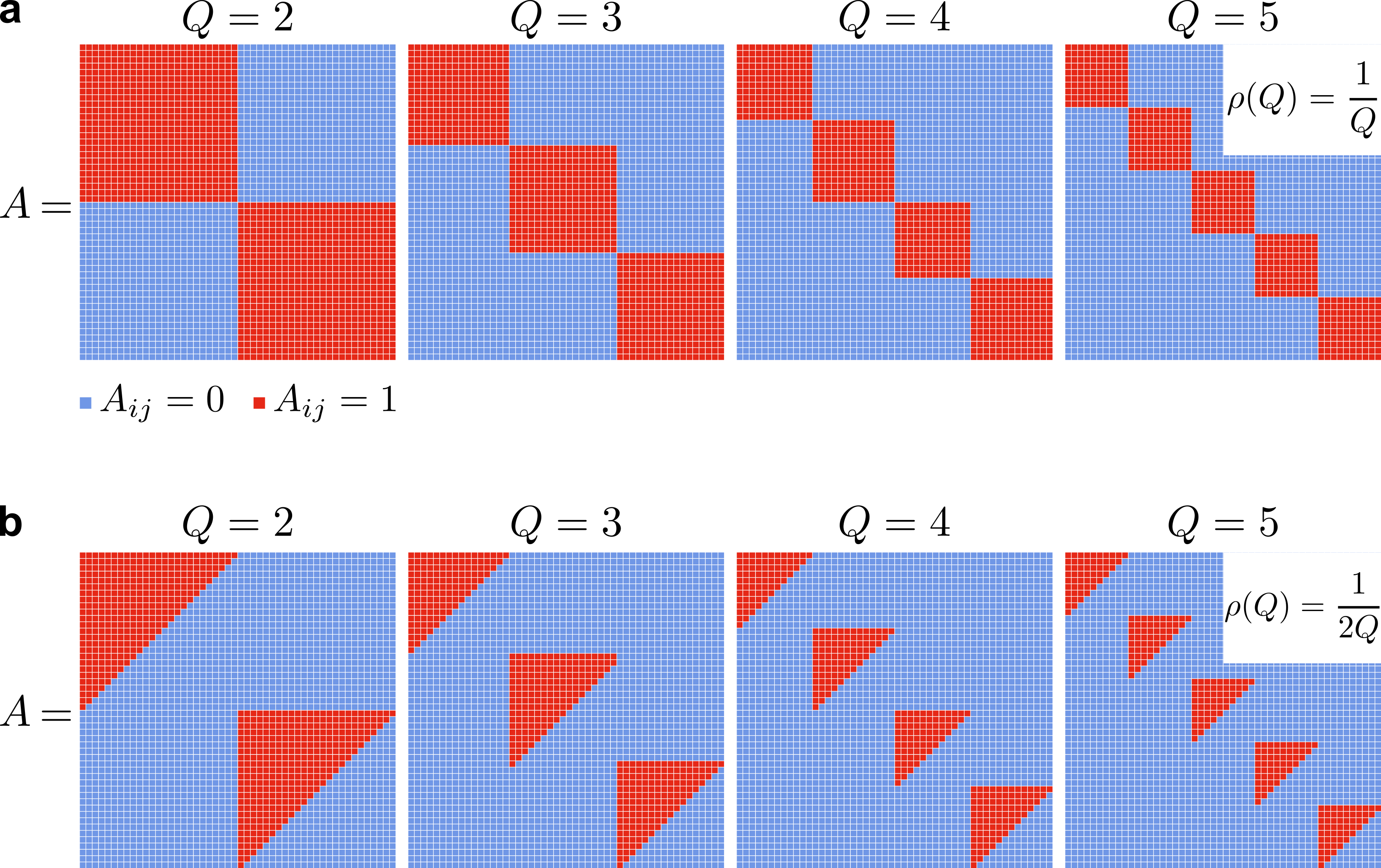}
\centering
\caption{{\bf Square filter and Triangle filter}. }
\label{fig:st_filter}
\end{figure}

The square filter $A$ is shown in Figure~\ref{fig:st_filter}a 
and is parameterized by a number $Q$ representing 
the number of blocks the matrix $A$ is divided into. 
The size of each block is $\frac{N}{Q}\times\frac{N}{Q}$. 
The $Q$ square blocks are arranged along the main diagonal. 
Mathematically, the entries $A_{ij}$ of $A$ are defined to 
be $0$ or $1$ by
\be
\begin{aligned}
A_{ij} &= 1\ \ \ {\rm if}\ \ \ 
\left\{
\begin{matrix}
i\in\left[1+(q-1)\frac{N}{Q}, q\frac{N}{Q}\right]\\
{\rm AND}\\
j\in\left[1+(q-1)\frac{N}{Q}, q\frac{N}{Q}\right]
\end{matrix}\ ,
\ \ \ q=1,\dots,Q
\right.\\
A_{ij} &= 0\ \ \ {\rm otherwise}
\end{aligned}
\ .
\label{eq:sfilter}
\ee
The density $\rho(Q)$ is easy to calculate and evaluates:
\be
\rho(Q) = \frac{1}{Q}\ .
\ee

%%%%%%%%%%%%%%%%%%%%
%%%%%%%%%%%%%%%%%%%%

\bigskip

The triangle filter is shown in Figure~\ref{fig:st_filter}b
and is parameterized by a number $Q$ representing 
the number of equal sized triangular blocks arranged 
along the main diagonal of the matrix. 
Mathematically, the entries $A_{ij}$ of $A$ are defined to 
be $0$ or $1$ by
\be
\begin{aligned}
A_{ij} &= 1\ \ \ {\rm if}\ \ \ 
\left\{
\begin{matrix}
i\in\left[1+(q-1)\frac{N}{Q}, q\frac{N}{Q}\right]\\
{\rm AND}\\
j\in\left[1+(q-1)\frac{N}{Q}, (2q-1)\frac{N}{Q}-i+1\right]
\end{matrix}\ ,
\ \ \ q=1,\dots,Q
\right.\\
A_{ij} &= 0\ \ \ {\rm otherwise}
\end{aligned}
\ .
\label{eq:tfilter}
\ee
The density $\rho(Q)$ of the triangle filter equals to
\be
\rho(Q) = \frac{1}{2Q}\ .
\ee

%HUB-INTERNEURON = 20
%INTERNEURON = 56
%INTERNEURON-MOTOR = 18
%INTERNEURON-SENSORY = 4
%INTERNEURON-SENSORY-MOTOR = 8
%MOTOR = 83
%SENSORY = 64

%My Machine: 2.3 GHz Intel Core i7

\end{document}